\documentstyle[prl,twocolumn,aps,floats,epsf,epsfig]{revtex}

\begin{document}

\draft
\twocolumn[\hsize\textwidth\columnwidth\hsize\csname @twocolumnfalse\endcsname

\title{On the temperature dependence of 2D ``metallic'' conductivity
  in Si inversion layers at intermediate temperatures}
\author{S. Das Sarma and E. H. Hwang}
\address{Condensed Matter Theory Center, 
Department of Physics, University of Maryland, College Park,
Maryland  20742-4111}

\maketitle

\begin{abstract}
We show that the recent experimental claim \{Pudalov {\it et
  al}. \prl {\bf 91}, 126403 (2003) \} of observing ``interaction
  effects in the 
conductivity of Si inversion layers at intermediate temperatures'' 
is incorrect and misleading. In particular, the temperature dependent
conductivity $\sigma$, in contrast to the resistivity (which is what
is shown in the paper), does not have a linear temperature regime,
rendering the extraction of the slope $d\sigma/dT$ completely
arbitrary. We also show that, at least for higher
densities, the standard semiclassical transport theory, which includes
realistic disorder effects such as scattering by {\it screened}
charged impurity and surface roughness, gives essentially quantitative
agreement with the experimental data.


\end{abstract}

\vspace{0.5cm}

]



In a recent Letter \cite{pudalov}, here after referred to as I,
Pudalov {\it et al}. have presented experimental results on the
temperature dependent resistivity $\rho(T)$ of Si inversion layers
comparing the data to the so-called interaction theory \cite{zala} of
Zala {\it et al}. Although claims of ``rigorous experimental test''
and ``excellent agreement'' are made rather uncritically in I, the
purpose of the current Comment is to point out the misleading
and essentially incorrect nature of the main claims in I. In
particular, we point out that (1) the comparison to the theory of
ref. \onlinecite{zala} carried out in I is inappropriate and
arbitrary; and (2) the well-established semiclassical Boltzmann
transport theory \cite{stern,ando,gold,dassarma} 
for Si inversion layers using
screened charged impurity scattering and interface roughness
scattering provides {\it quantitative} agreement with much of the
higher density data ($n \ge 5 \times 10^{11} cm^{-2}$) presented in I
with the agreement becoming qualitative at lower densities.

\begin{figure}
\centerline{\epsfig{figure=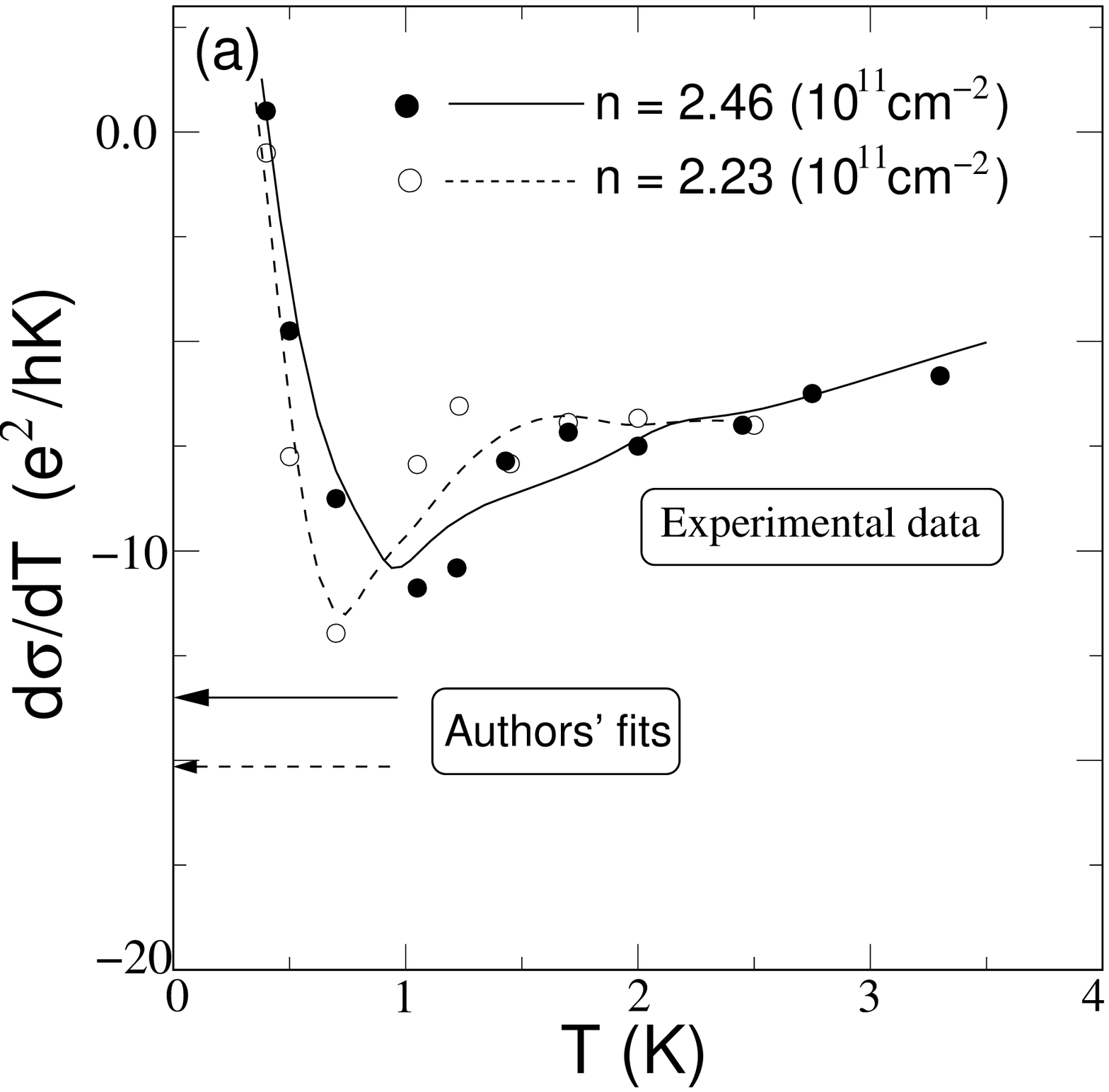,width=2.8in}}
\vspace{0.5cm}
\centerline{\epsfig{figure=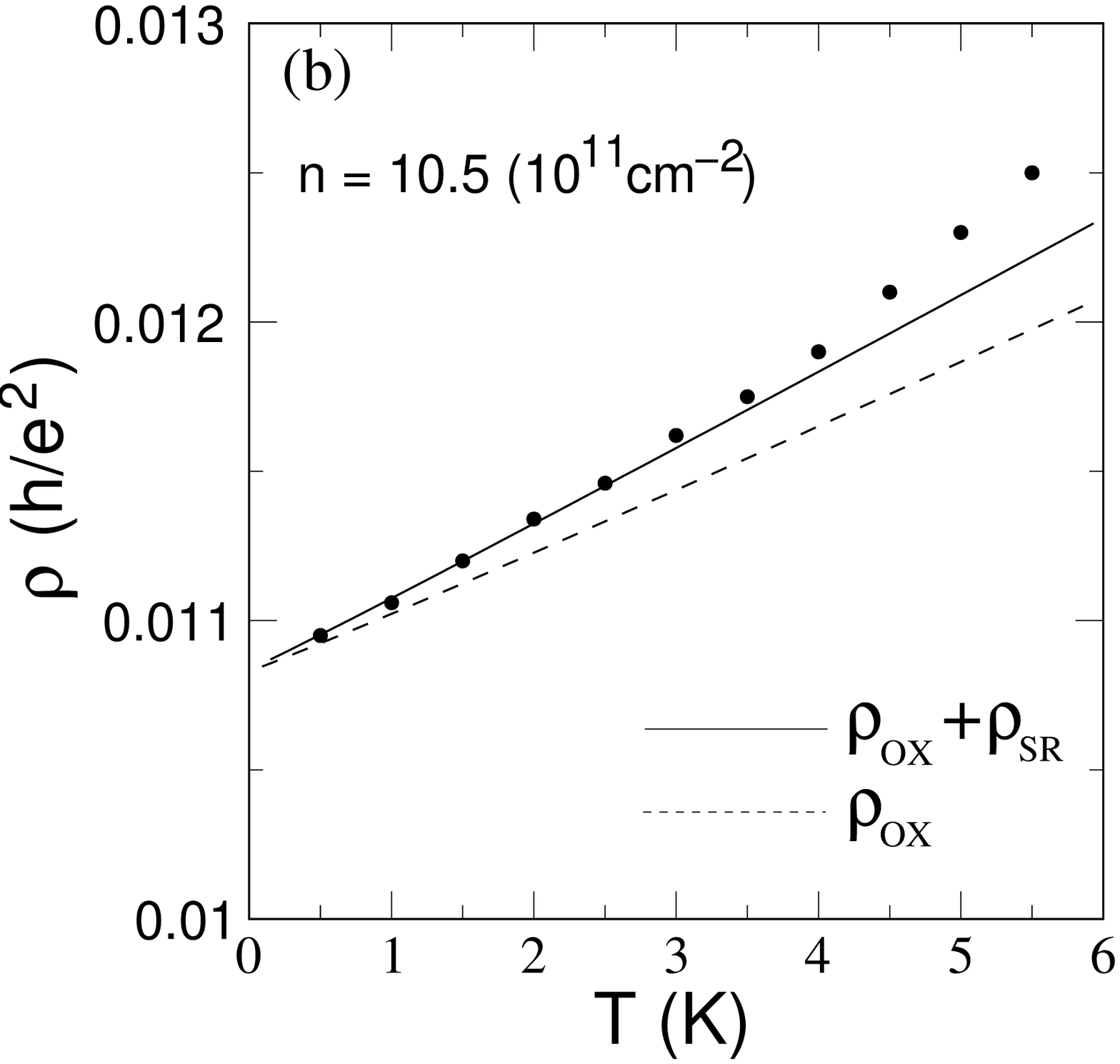,width=3.in}}
\caption{(a) $d\sigma/dT$ using an interpolation scheme (lines)
for the data of I with the slope used in I shown as
arrows. Dots in (a) represent the average slope at the data points
without any interpolation.
(b) The calculated resistivity 
including screened interface charged impurity and surface roughness
scattering. Dots indicate experimental data from I.}
\end{figure}

The interaction theory \cite{zala}, with which Pudalov {\it et al.}
compare their data, predicts a linear T-dependence of the
conductivity $\sigma(T)/\sigma_0 = 1 + F \frac{T}{T_F}$ 
in the intermediate temperature ballistic regime 
where $\sigma_0$ is the $T=0$ ``Drude'' conductivity 
and $T_F$ the Fermi temperature. The slope 
$F \equiv F(r_s)$, 
which depends on the dimensionless
$r_s$ ($\propto n^{-1/2}$) parameter, cannot be calculated within the
theory and is predicted to change its sign at high density (which has
not yet been observed experimentally). The absolute necessary
condition (but, by no means sufficient) for verifying the theory of
ref. \onlinecite{zala} is obviously an observation of a linear
temperature dependent conductivity over a reasonable temperature range.
This is particularly
so in view of the fact that the interaction theory \cite{zala} can
only predict the qualitative leading order temperature dependence, but
{\it not} the quantitative magnitude of the slope $F$ for the
$\sigma(T)$ curves. A rudimentary analysis of the data presented in I
clearly shows that $\sigma(T)$ results of I are nonlinear, and thus do
not satisfy the minimal necessary condition needed for a comparison
with the interaction theory. The authors of I have taken the {\it
misleading} step of presenting their data for the resistivity
$\rho(T)$, rather than the conductivity
$\sigma(T)=[\rho(T)]^{-1}$. The $\rho(T)$ data in I superficially
appear relatively more linear than the corresponding $\sigma(T)$
results, and quite trivially even a linear $\rho(T)$
is {\it not} equivalent to a linear
$\sigma(T)$ unless the temperature correction is small,
which is manifestly not the case in I ({\it i.e.}
the temperature dependence in not weak except at the highest densities
where the Boltzmann theory gives quantitatively accurate results).

We show in Fig. 1(a) our best estimate for $d\sigma/dT$ extracted
numerically for the data of I, and it is obvious that a linear-T
approximation to $\sigma(T)$, {\it i.e.} a constant $d\sigma/dT$ over
any reasonable range of temperature, simply does not apply anywhere in
the data making the whole exercise of the comparison to the
interaction theory completely arbitrary.
In Fig. 1(b) 
we show an essentially parameter-free
quantitative comparison between the high-density data in I and the
standard semiclassical Boltzmann transport theory including realistic
effects of screened interface charged impurity and surface roughness
scattering. This quantitative agreement remains good down to about
$5\times 10^{11}cm^{-2}$ below which the experimental $\rho(T)$ shows
stronger nonlinear temperature dependence at higher ($>3K$)
temperature, but the intermediate temperature ($\sim 1K$) slope
$d\sigma/dT$ continues to agree well with the prediction of the
realistic Boltzmann theory. The quantitative agreement between the
Boltzmann theory and the data of I can be further improved (down to lower
densities) by including scattering by bulk charged impurity centers
and by adjusting the carrier density and/or carrier effective mass (as
well as an improved screening function including effects of scattering
and local field correction).

We conclude by pointing out that the $\rho(T)$ data presented in I
shows smooth evolution from high to low density with only the
quantitative temperature dependence becoming stronger in a continuous
manner with decreasing carrier density. This indicates that the
temperature dependent disorder (e.g. screened Coulomb scattering) is
playing an important role since at high and intermediate densities one
gets quantitative agreement between the realistic Boltzmann theory and
the experimental $\rho(T)$ --- qualitatively, lowering density is
weakening screening leading to stronger temperature dependent disorder
and consequently larger resistivity.

\end{document}